\documentclass[preprint,aps,prfluids,superscriptaddress]{revtex4-2}
\usepackage[utf8]{inputenc}
\usepackage{amsmath}
\usepackage{amsfonts}
\usepackage{amssymb}
\usepackage{graphicx}
\usepackage{float}
\usepackage{siunitx}
\usepackage{hyperref}

\begin{document}

\title{Probing interfacial effects with thermocapillary flows}
\author{M. Maza-Cuello}
\affiliation{CNRS, Sciences et Ing\'enierie de la Mati\`ere Molle, ESPCI Paris, PSL Research University, Sorbonne Universit\'e, 75005 Paris, France}
\author{C. Fr\'etigny}
\affiliation{CNRS, Sciences et Ing\'enierie de la Mati\`ere Molle, ESPCI Paris, PSL Research University, Sorbonne Universit\'e, 75005 Paris, France}
\author{L. Talini}
\affiliation{CNRS, Surface du Verre et Interfaces, Saint-Gobain, 93300 Aubervilliers, France}
\date{\today}

\begin{abstract}
We report on the thinning of supported liquid films driven by thermocapillarity. The liquids are oil films, of initial thicknesses of a few tens of microns. A local and moderate heating of the glass substrate on which they are spread on induces a thermocapillary flow, which allows the formation of ultra-thin films. We show that, within a given time range, the thinning dynamics of submicron thick films is governed by the thermocapillary stress. It results in a simple dependency of the thickness profiles with time, which is evidenced by the collapse of the data onto a master curve. The master curve only depends on the liquid properties and on the thermal gradient, and allows a measurement of the latter. As the films further thin down to thicknesses within the range of molecular interactions, a deviation from the master curve appears. Although all investigated oils are supposedly fully wetting glass, the nature of the deviation differs between alkanes and silicone oils. We attribute the observed behaviors to differing signs of disjoining pressures, and this picture is confirmed by numerical resolution of the thin-film equation. We suggest thermocapillary flows can be used to finely probe molecular interactions.
\end{abstract}

\maketitle

\section{Introduction \label{s:intro}}

The surface tension of a pure liquid decreases with increasing temperature and a temperature gradient at its free surface results in a thermocapillary stress, inducing a flow away from the hottest region. The migration of liquids induced by a temperature gradient has been known and exploited for a long time \cite{hershey_ridges_1939}. It is used for instance for the patterning of surfaces, in the context of micromachining with molten metals \cite{anthony_surface_1977, willis_transport_2000}. In this process, a strong laser beam is shined on the surface of a body of metal, producing the melting of an upper layer which is driven mainly by the surface tension gradient. In recent years, a similar principle has been applied in the micropatterning of polymer melts for lithography techniques \cite{singer_thermocapillary_2017}. In contrast, in other applications thermocapillary flows may be undesired. For instance, they are known to cause failure of lubrication in tribo-systems in which thermal gradients spontaneously appear; therefore, the design of devices must be thought in consequence \cite{tribo_systems}.
From a more fundamental point of view, the behavior of a supported liquid film submitted to thermal gradients has been  studied in different configurations. Theoretical works have focused on the dimpling of the free surface of the film that results from a thermocapillary stress, and to which both capillarity and gravity oppose \cite{hershey_ridges_1939, tan_steady_1990, pimputkar_transient_1980, yeo_marangoni_2003}. Experimentally, controlled thermal gradients at the free surface can be created by locally heating the liquid by radiative transfer \cite{hitt_radiationdriven_1993}, but more usually it is the substrate they lie on that is heated. Focusing a laser at an absorbing substrate provides an easy way to locally increase its temperature and create a temperature gradient at a liquid interface by heat transfer \cite{chraibi_thermocapillary_2012, wedershoven_infrared_2014, pottier_prl_2015, klyuev_thermocapillary_2021}. Whatever the effect inducing the thermocapillary stress, the latter allows the formation and control of thin liquid films, in a region whose size is given by the extent of the thermal gradient \cite{tan_steady_1990}. Within this region, in which the films can reach submicron thicknesses even with a moderate temperature rise, intermolecular forces have non negligible effects, which in particular condition the rupture of the film. Thermocapillary-induced dewetting of liquids on solid substrates has thus been investigated by monitoring the thickness at the central hottest point \cite{wedershoven_infrared_2014, klyuev_thermocapillary_2021}. However, up to date, poor attention has been paid to the dynamics of the whole thin film region, in which intermolecular forces are large. Here, we show that interfacial effects at very small scales can be probed through accurate measurements of thickness variations in this region.

The paper is organized as follows. We describe the experimental setup and methods in section \ref{s:experiments}.
A review of the hypotheses that are made leading to a well-known thin film equation describing the evolution of a thin liquid film under the action of thermocapillarity are discussed in section \ref{s:theory}. We further present the experimental results in section \ref{s:results}. Finally, section \ref{s:conclusion} provides a conclusion and outlook for future work.

\section{Experimental set-up and methods \label{s:experiments}}
\subsection{Experimental set-up}
\begin{figure}[h!]
	\centering
	\includegraphics[width=0.9\linewidth]{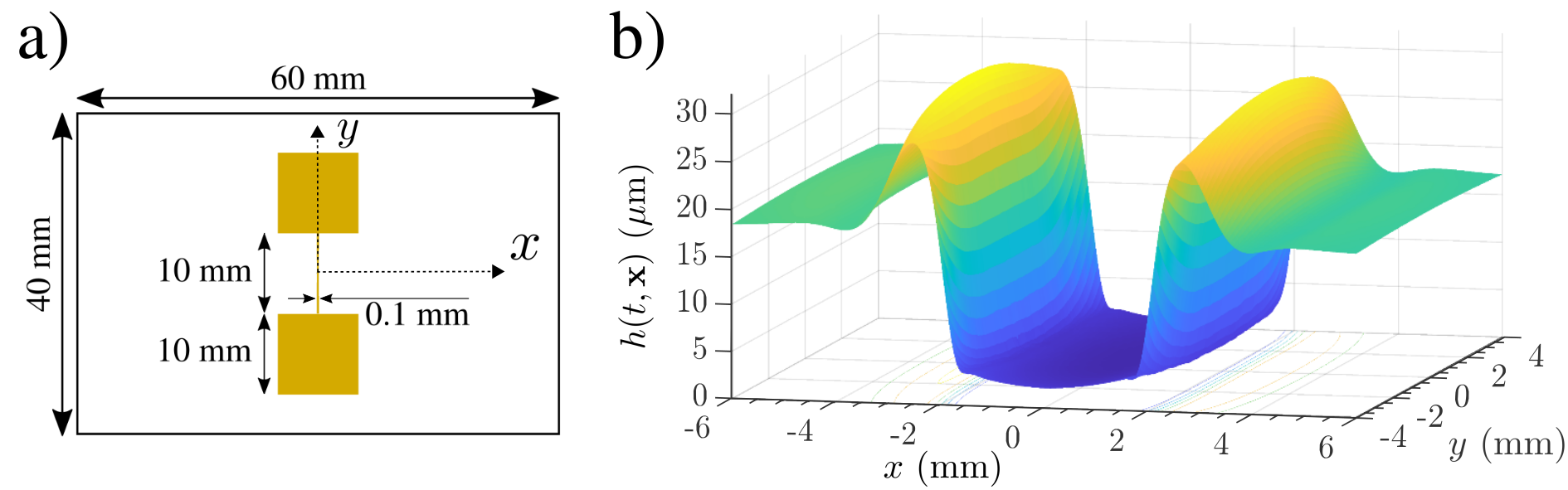}
	\caption{\label{fig:slide_schlieren}a) Top view of the glass substrate. On its bottom side, a microprinted circuit connected to a voltage generator creates the thermal gradient. 
	b) Typical thickness surface $h(t,\mathbf{x})$ at an intermediate stage of the dynamics, reconstructed surface via Free-surface synthetic Schlieren method \cite{moisy_synthetic_2009}. The thermocapillary effect promotes the apparition of a very thin ($h < \SI{1}{\micro\metre}$) layer that extends over several millimetres.}
\end{figure}

The liquid films are spread by spin-coating on substrates that are rectangular glass slides of thickness \SI{100}{\micro\metre}. 
Prior to deposition, the top side of the substrate is washed by spin-coating isopropanol and acetone over it several times, letting it dry finally for a couple of minutes before spreading the liquid. 
An electrical circuit is printed on the bottom side of the slides, with a pattern consisting in two squares connected by a line, as shown in Fig.~\ref{fig:slide_schlieren}a). The squares are connected to a dc voltage generator and the resulting electrical current is measured. The central line, of width \SI{100}{\micro\meter}, acts as a heat dissipator, generating a thermal field. We denote $O$ the centre of the line and $Ox$ and $Oy$ the axes respectively perpendicular and parallel to the line. We make the approximation that the induced thermal field does not depend on $y$, which is justified in the vicinity of the centre of the line. We define the thermal excess $\theta(x) = T(x) - T_0$ where $T_0$ is the room temperature and $T(x)$ the thermal field.
The dissipated power $P$ along the line is a control parameter that sets the maximal temperature rise, $\theta_{max}=\theta(0)$. In our experiments $P \sim \SI{1}{\milli\watt}$ and $\theta_{max}\sim\SI{1}{K}$. Infra-red measurements have evidenced that the horizontal extent of the thermal gradient $w$ is a few millimeters. 
The thermal gradient drives a flow in the liquid films away from the central line, inducing a thinning of the films along that line. 
A typical surface of a film once a very thin film has been formed at the central region is shown in Fig.~\ref{fig:slide_schlieren}b). The profile was measured via an implementation of the Free-surface synthetic Schlieren method \cite{moisy_synthetic_2009}, which we will not detail here since we focus on measurements at smaller thicknesses, which are described in the following. The region we focus on  is that at the centre (where the heating line lies), which has been emptied of liquid so that the thickness is below the micrometre scale in an area spanning several millimetres. We emphasise that, as expected, the measured profile only very weakly depends on $y$, which justifies the 1D profile we will consider in the following. 
Experiments were performed with different initial film thicknesses, $h_0$ and electrical powers $P$. The initial thickness $h_0$ is measured by optical profilometry just after spreading the films.

\subsection{Investigated liquids}
The investigated liquids are silicone oil (ABCR AB112154), $n$-hexadecane (Sigma-Aldrich) and a ramified isomer of hexadecane: 2,2,4,4,6,8,8-heptamethylnonane (HMN, Sigma-Aldrich). Their useful physical properties are listed in table \ref{t:physical_props}; in particular, they have very close values of coefficient of surface tension variation, $\gamma_\theta = -\partial_T\gamma|_{T = T_0}$. In contrast, we emphasise that their melting temperatures are very different.

Since the temperature increase imposed to produce the thermocapillary effect remains modest, the variations of viscosity and density are small.
An estimation from the thermal dependencies of these properties for the three liquids investigated \cite{roberts_physical_2017, dubey_temperature_2008, luning_prak_density_2014} gives a variation on viscosity and density of around 2~\% and 0.1~\%, respectively, for a thermal rise of $\theta_{max} = \SI{1}{\kelvin}$.
Similarly, the variation of refractive index of alkanes \cite{camin_physical_1954} and silicone oil \cite{van_raalte_measurement_1960} caused by this temperature change is at most $0.01$~\%.
Therefore, the variations of viscosity, density and refractive index produced by heating the liquid will be neglected in the following.

\begin{table}[h!]
    \centering
    \begin{ruledtabular}
\begin{tabular}{crrrrl}
  & $\eta$ ($\si{\milli\pascal\second}$) & $\rho$ ($\si{\kilo\gram\per\cubic\metre}$)& $\gamma_\theta$ ($\si{\newton\per\metre\per\kelvin}$) & $T_{\text{melt}}$ ($\si{\celsius}$) & ref.\\
  \hline
Silicone oil & $19$ & $950$ & $6.5\times 10^{-5}$ & $< -40$ &  \cite{ abcr20},\cite{ricci_density_1986}\\
$n$-C$_{16}$ & $2.2$ & $733$ & $8.7\times 10^{-5}$  & $18$ & \cite{korosi_density_1981},\cite{dubey_temperature_2008},\cite{ocko_surface_1997} \\
HMN & $3.2$ & $781$ & $8.7\times 10^{-5}$ & $-4$ &\cite{korosi_density_1981},\cite{luning_prak_density_2014},\cite{HMN_Tm}
\end{tabular}
\end{ruledtabular}
\caption{\label{t:physical_props}Investigated liquids with their main properties: viscosity, $\eta$, density, $\rho$, thermal coefficient of the surface tension, $\gamma_\theta = -\partial_T\gamma|_{T = T_0}$, and melting temperature, $T_{\text{melt}}$.}
\end{table}

\subsection{Thickness measurements}

\begin{figure}[h!]
	\centering
	\includegraphics[width=0.9\linewidth]{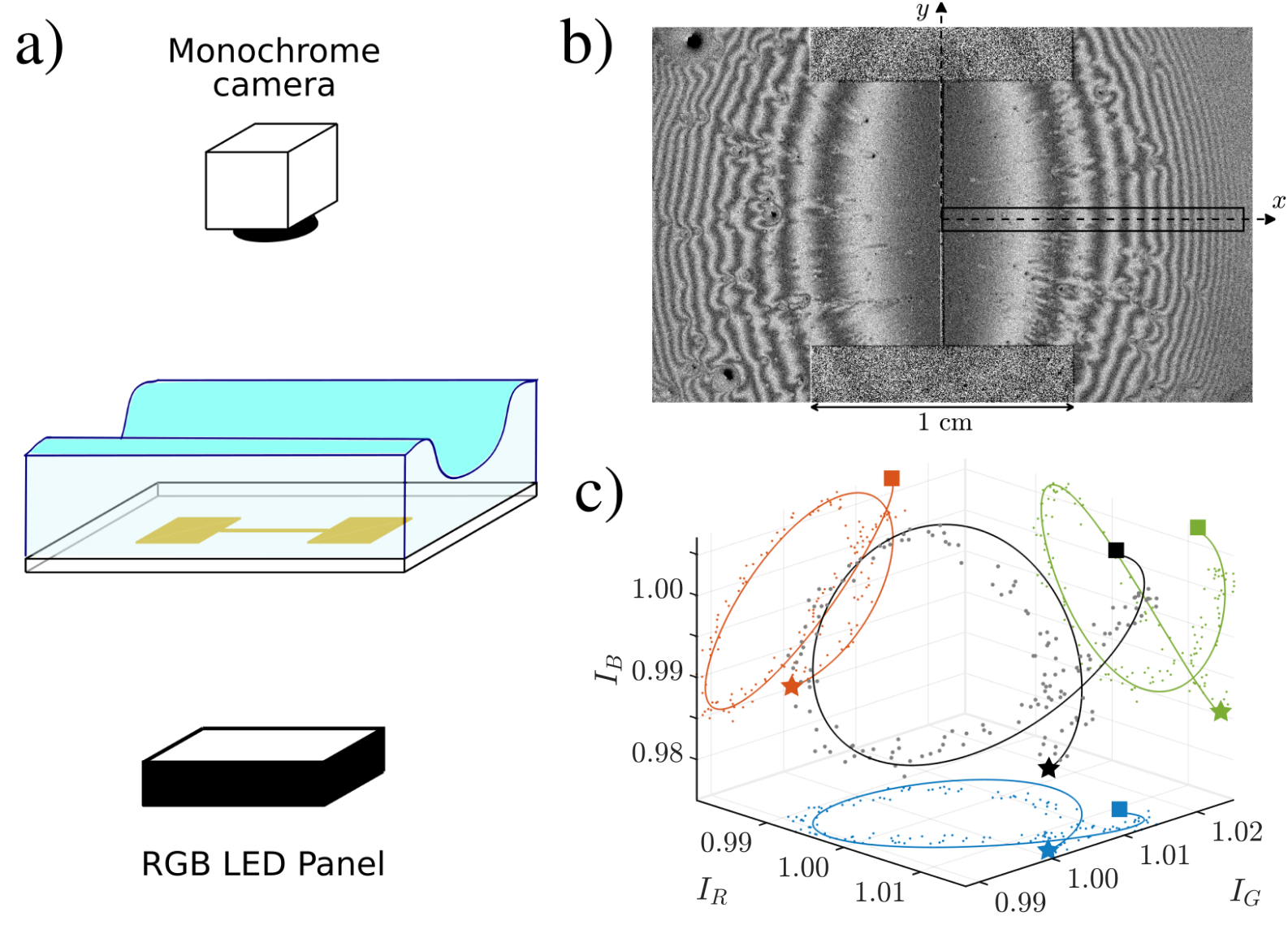}
	\caption{\label{fig:interferometry}a) Sketch of the measurement setup. A RGB Panel illuminates from below the experimental cell, consisting of a glass substrate over which a liquid film is deposited via spin-coating. A monochrome camera on top takes pictures at specified intervals. 
	b) Typical interferometric profile of a silicone oil film obtained after an initial transient period. Here $\lambda_R = \SI{635}{\nano\metre}$ (red) and the height different between intensity maxima is $\Delta h = \SI{113}{\nano\metre}$. The black rectangle indicates the region of interest for measuring the film dynamics, the interference intensity $I_R(x)$ is obtained by performing an average along the vertical direction (and similarly for green, $I_G$ and blue, $I_B$).
	c) Grey dots: interferometric data in intensity space $(I_R, I_G, I_B)$ for the film at a given time. Black line: theoretical intensity curve, with the thickness $h$ as parameter. The coloured points and lines are projections into the orthogonal planes for better visualization. Fitting the data points to the theoretical curve gives an absolute measurement of the thickness $h$ at the corresponding $x$ coordinate. The squares corresponds to $h = \SI{80}{\nano\metre}$, the stars to $h = \SI{300}{\nano\metre}$.}
\end{figure}

The thickness of the film at the central region, where it will be below the micrometre scale, is measured with 3-colour interferometry \cite{schilling_absolute_2004, de_ruiter_air_2015}.
The measurement setup is sketched in Fig.~\ref{fig:interferometry}a).
The cell is shined from below using a RGB-LED of peak wavelengths $\lambda_R = \SI{635}{\nano\metre}$, $\lambda_G = \SI{518}{\nano\metre}$, and $\lambda_B = \SI{463}{\nano\metre}$.
Images are taken from above with a CMOS monochrome camera.
Images were taken typically each \SI{10}{\minute} for silicone oil during a couple of days, and each \SI{10}{\second} during some hours for alkanes.

At any given measuring time, three images, one for each colour is taken.
This leads to interference patterns such as the one shown in Fig.~\ref{fig:interferometry}b).
From such an image, an interference profile $I_C(x)$ is obtained at each time and for each colour $C = R,\,G,\,B$ by averaging the intensity profiles along the $y$-axis within a region spaning around \SI{1}{\milli\metre}, which corresponds to the rectangle in Fig.~\ref{fig:interferometry}b).
To obtain the thickness $h(x)$ from $I_C(x)$, one considers the theoretical linearised expression of transmission interference through a thin film of thickness $h$,
\begin{equation}
	I_C^{th}(h) \propto 1 - A\cos\left(\dfrac{4\pi n h}{\lambda_C}\right) 
\end{equation}
where $n$ is the refractive index of the liquid and $A$ is a Fresnel coefficient depending on the refractive indexes of the media considered, in our experiments $A \simeq 0.012$.
In the computations, the interference profiles $I_C(h)$ are convolved with a Gaussian function mimicking the broad-band emission of the LED source for each channel.

Working on the intensity space $\mathbf{I} = (I_R, I_G, I_B)$, as shown in Fig.~\ref{fig:interferometry}c), it is possible to use $h$ as a fitting parameter of the theoretical curve for the data points, resulting in an absolute-thickness curve $h(x)$ at the time of measurement.
This procedure allows to obtain thicknesses below the first-order fringe, i.e. smaller than \SI{100}{\nano\metre}, thus in the range where intermolecular forces are to be expected. In our current implementation, overall thickness profiles ranging from the tens of nanometres to the micron scale can be measured with a nanometric accuracy. The floor sensitivity of the measurement is limited by experimental noise, which mainly originates from the camera sensor; as a result, thicknesses below $10$~nm cannot be determined.
Due to the opacity of the heating line, the film thickness cannot be measured for $|x| \lesssim \SI{0.1}{\milli\metre}$.

\subsection{Numerical integration}
Numerical integration of the thin film equation discussed below is done using the \texttt{pdepe} numerical solver of \textsc{Matlab} \cite{*[{See }] [{ for the documentation page of \texttt{pdepe}. Run with \textsc{Matlab} R2022a.}] matlab_pdepe}. Since the equation is fourth order on the thickness gradient, the second derivative $u(t,x) \equiv \partial_x^2h$ is considered as an independent field and a simple closure expression $u - \partial_x\left\{ \partial_x h\right\} = 0$ is simultaneously solved. 
The initial condition is a film of homogeneous thickness, $h(0,x) = h_0$, $u(0,x) = 0$.
For ease of convergence, the $x$ spacing is logarithmic, allowing to better describe the evolution close to the centre as well as capturing the long range evolution.
The numerical integration typically lasted some hours.

\section{Theoretical background \label{s:theory}}
As schemed in figure \ref{fig:variables}, we consider a homogeneous layer of incompressible liquid of constant viscosity $\eta$ and density $\rho$ with initial thickness $h_0$, that is submitted to a symmetric temperature excess $\theta(x)$. The excess temperature remains small:  $\theta_{max}=\theta(0)\ll T_0$, with $T_0$ the room temperature and all the properties of the liquid are assumed to be uniform. The resulting dynamics is assumed to be invariant along the $y$-axis. In addition, since we use poorly volatile fluids we consider that evaporation is negligible.

\subsection{Hypotheses and dimensionless numbers}

The order of magnitude of the thermocapillary stress is $\gamma_{\theta} \theta_{max}/w\sim\SI{0.1}{\pascal}$. The resulting velocity in a liquid film of thickness \SI{10}{\micro\metre} and viscosity $10^{-2}$~\si{\pascal\second} is $v \sim \SI{100}{\micro\metre\per\second}$.
Since the liquids considered in the experiments have a thermal diffusivity $\kappa \sim 10^{-7}\si{\metre\squared\per\second}$ \cite{rhodorsil, velez_temperature-dependent_2015}, the thermal Péclet number $\text{Pe} = h_0v/\kappa \sim 10^{-2}$ is small enough to consider thermal convection negligible. Therefore, the transport of heat from the substrate to the liquid is purely diffusive. In addition, the thermal field is established in a time $h_0^2/\kappa \sim 10^{-2}$~\si{\second}, much faster than the timescale of hours involved in the dynamics. Hence, we assume that the thermal field is established instantaneously and that it is constant throughout the film's thickness.
The Reynolds number $\text{Re} = h_0v/\eta \sim 10^{-7}$ is also very small.

In the literature, a dimensionless number has been defined to measure the effect of the thermocapillary stress compared with that of hydrostatic pressure, $R = 3\Delta\gamma / 2\rho gh_0^2$, where $\Delta\gamma = \gamma_\theta\theta_{max}$ \cite{tan_steady_1990}. We typically work with $R \sim 10$, a larger thermocapillary effect than in Ref. \cite{burelbach_steady_1990} so that the thickness at the centre will always reach the region where intermolecular effects are relevant \cite{tan_steady_1990}.


\subsection{Thin-film equation}
In experiments, the thermal field decays over a typical length $w \sim \SI{2}{\milli\metre}$ much larger than the initial thickness, $h_0 \ll w$, allowing to consider a small gradient approximation. 
As a result, it is possible to obtain a lubrication approximation of the Navier-Stokes equations leading to a partial differential equation for the thickness profile $h(t,x)$ belonging to the family of thin film equations \cite{oron_long-scale_1997}.

\begin{figure}[h!]
	\centering
	\includegraphics[width=0.5\linewidth]{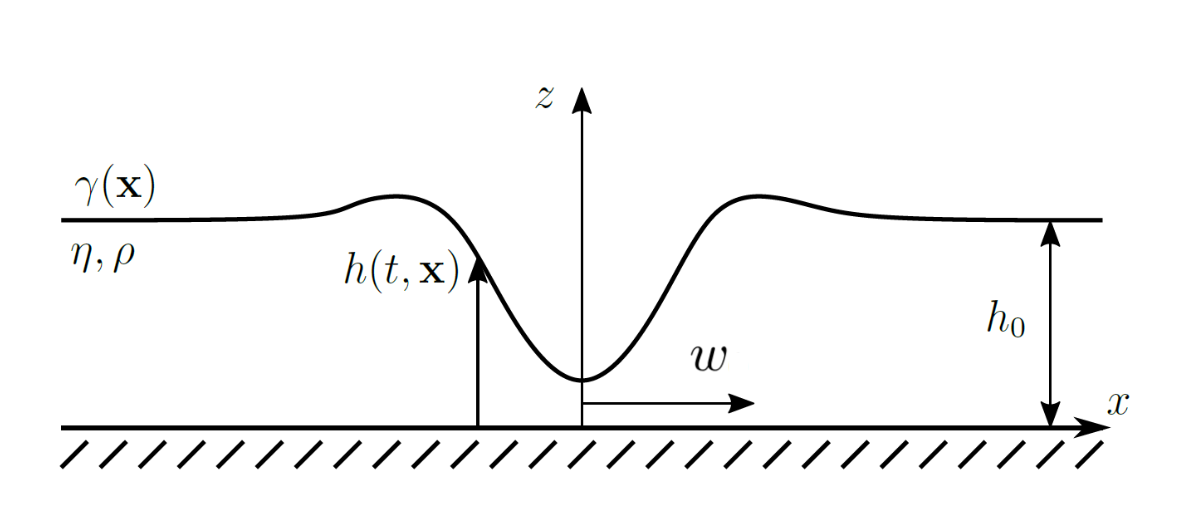}
	\caption{\label{fig:variables}Sketch of the film dynamics with main variables: a thin liquid film of density $\rho$ and viscosity $\eta$, with initial thickness $h_0$, is deposited on top of a glass substrate which is heated locally, $w$ being the decay length of the thermal field.}
\end{figure}

For the system defined by the conditions above and schematized in Fig.~\ref{fig:variables}, the thin film equation reads \cite{tan_steady_1990}
\begin{equation}
	\partial_th 
	+
	\dfrac{1}{3\eta}\partial_x
	\left\{
		h^3\partial_x
		\left[
			\gamma_0\partial_x^2h - \rho g h + \Pi(h)
		\right]
	\right\}
	-
	\dfrac{\gamma_\theta}{2\eta}\partial_x
	\left\{
		h^2\partial_x\theta(x)
	\right\}
	=
	0,
	\label{eq:TFE}
\end{equation}
where the pressure contributions considered on the second term of the l.h.s. are the capillary pressure, the hydrostatic pressure and $\Pi(h)$ is a disjoining pressure term accounting for the intermolecular interactions.
The third term represents the thermocapillary force driving film thinning. Capillarity and gravity oppose film thinning whereas disjoining pressure may oppose or amplify thinning, according to its sign. The relative importance of the pressure terms varies as the film thins down and this variation has been investigated theoretically \cite{burelbach_steady_1990}. If the initial thickness of the film is larger than the range of molecular interactions, then in a first stage, the film dimples and both gravity and capillarity contribute to slow down its evolution. The central region of the film further empties and a central area of very small thickness is formed. In this region of small thickness and small curvature, both capillary and hydrostatic pressures are negligible. As a result, film thinning results from an interplay of thermocapillary stress and disjoining pressure. If the intermolecular forces are attractive then spontaneous rupture of the film occurs. In the case of repulsive intermolecular forces, a fully equilibrated stationary state can be reached and has been observed in experiments \cite{tan_steady_1990,clavaud_modification_2021}; it corresponds to a balance of the thermocapillary stress and disjoining pressure. In such a state, it would be possible, in principle, for the disjoining pressure to be measured if the thickness of the ultra-thin liquid film at the central region is measured and the imposed thermal field is known. However, most experimental works have not aimed at measuring disjoining pressure and rather focused on the question of the film rupture \cite{burelbach_steady_1990, wedershoven_infrared_2014}. In addition, an accurate determination of the thermal field is required to measure the disjoining pressure, which is a challenging task.
In the geometry we investigate, the thermal gradients extend over millimetric lengths and we observe a very slow dynamics of the central region, which thins down during a time scale of days. In practice, this implies that the full equilibrium cannot be experimentally reached since other effects, such as sample contamination or liquid evaporation, appear at these very long time scales. Nevertheless, the thinning dynamics itself provides information on the intermolecular forces. Therefore, rather than trying to reach a stationary state, we take profit of the thinning dynamics to probe the intermolecular interactions at stake.

\subsection{Pure thermocapillary regime}
Before focusing on the effects of disjoining pressure, we first examine the preceding stage, in which the central area of small thickness has formed but its thickness is large enough for disjoining pressure to be neglected, i.e. larger than roughly $100$~nm. In this stage, the capillary and gravitational pressures can as well be neglected in the description of the evolution, consistently with former works \cite{tan_steady_1990, burelbach_steady_1990}.
Film thinning is then entirely driven by thermocapillarity, which no effect opposes. Equation (\ref{eq:TFE}) then becomes
\begin{equation}
	\partial_th
	-
	\dfrac{\gamma_\theta}{2\eta}
	\partial_x
	\left\{
		h^2\partial_x\theta(x)
	\right\}
	=
	0.
	\label{eq:pureTC}
\end{equation}

Equation (\ref{eq:pureTC}) can be solved by separation of variables, giving
\begin{equation}
	h(t,x) 
	= 
	\dfrac{h_0}{1+(t/t^*)}
	\dfrac{|\partial^2_x\theta(0)|}{2\sqrt{|\partial_x\theta(x)|}}
	\int_{0}^{x}\dfrac{ds}{\sqrt{|\partial_s\theta(s)|}},
	\label{eq:hTC}
\end{equation}
where $t^* = 2\eta/(\gamma_\theta|\partial^2_x\theta(0)|h_0)$.
Since $t^*$ is at most a few seconds and the timescale of the experiments is hours, the limit $t \gg t^*$ is quickly reached. Equation (\ref{eq:hTC}) then simplifies and the product $t\times h(t,x)$ can be expressed as 
\begin{equation}
	t\times h(t,x) 
	= 
	\dfrac{\eta}{\gamma_\theta}
	\dfrac{1}{\sqrt{|\partial_x\theta(x)|}}
	\int_{0}^{x}\dfrac{ds}{\sqrt{|\partial_s\theta(s)|}}.
	\label{eq:mastercurve}
\end{equation}
The product $t\times h(t,x)$ is therefore independent of time and solely a function of the spatial coordinate, $x$. Since in addition it only depends on the liquid properties and not on the initial film thickness, for a given liquid, $t\times h(t,x)$ curves are expected to collapse onto a master curve that is a function of $x$ only, whatever the initial conditions of the experiment.
Inversely, at any given location $x$, the thickness has a simple dependency on time: $h(t, x) \propto t^{-1}$. This temporal scaling has been reported in experiments performed by heating a substrate with a laser beam \cite{wedershoven_infrared_2014}. The variations with time of the thickness of the oil film spread on the substrate were found to follow $h(t, 0) \propto t^{-1}$ for a wide range of laser power and beam waists under half a micrometre. It confirms the relevancy of a regime in which both capillarity and gravity are negligible. In the former work, the role of capillarity was found to be apparent only for very small beam waists at and below the micrometre scale, where the curvature of the film is sufficiently large. We demonstrate in the following that, in our experiments, capillarity and gravity are always negligible in the central area. 

\section{Experimental results \label{s:results}}

\subsection{Master evolution at intermediate thicknesses \label{sb:master_curve}}
\begin{figure}[h!]
	\centering
	\includegraphics[width=\linewidth]{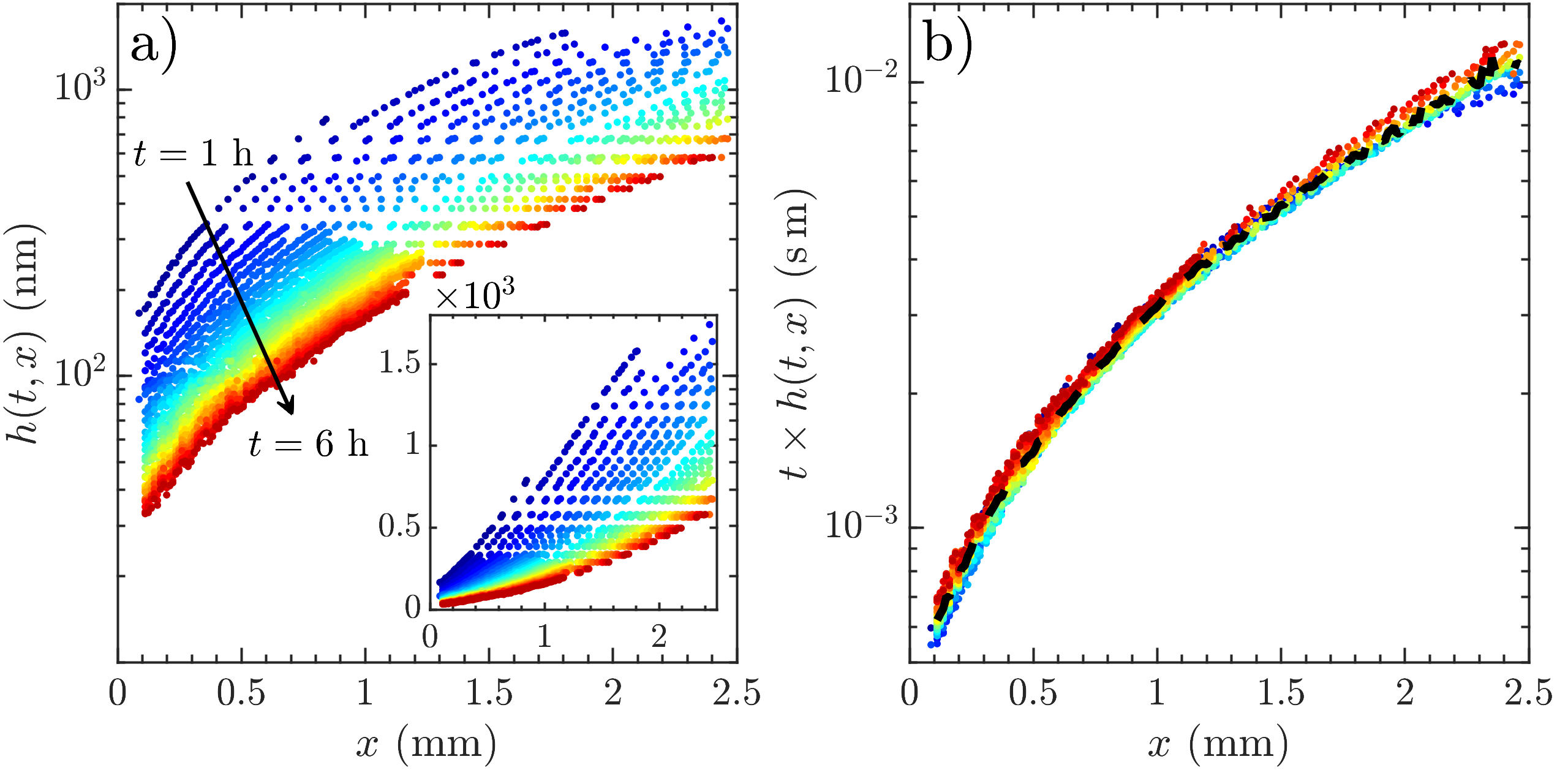}
	\caption{\label{fig:master_curve_silicone}a) Experimental profiles of an experiment with silicone oil with $h_0 = \SI{36\pm2}{\micro\metre}$ and $P = \SI{0.81\pm0.01}{\milli\watt}$.
	The time between profiles is \SI{10}{\minute}.
	The inset shows the same profiles in linear scale. 
	b) The products $t \times h(t,x)$ for the experimental profiles of panel a) at corresponding times collapse onto a master curve, which indicates that the local dynamics is solely described by the thermocapillary effect. The dashed line is the average of the experimental curves that is used to determine the thermal gradient.}
\end{figure}

In figure~\ref{fig:master_curve_silicone}, we show the result of a typical measurement of the evolution at the central area, extending roughly a few millimeters away from the central line, similar as the one in figure \ref{fig:slide_schlieren}b). The evolution is displayed, between 1 and 6 hours after turning the thermal gradient on. Since the evolution is symmetric with respect to the origin of the $x$-axis, the thickness of only one side of the film is reported. The experiment uses a film of silicone oil with initial thickness $h_0 = \SI{36\pm2}{\micro\metre}$ and the dissipated power is $P = \SI{0.81\pm0.01}{\milli\watt}$.

The film thickness increases very slowly from the centre to the edge: it reaches a few micrometers only over a distance of a few millimeters. The spatial variations of $t \times h(t,x)$ are  also displayed in the figures: remarkably, all curves nicely collapse onto a master curve. According to the analysis of section \ref{s:theory}, the collapse indicates that the effects of both capillary and hydrostatic pressures are negligible and that the dynamics is driven by thermocapillarity only, following the simple time dependency given by equation (\ref{eq:mastercurve}).

\begin{figure}[h!]
	\centering
	\includegraphics[width=0.5\linewidth]{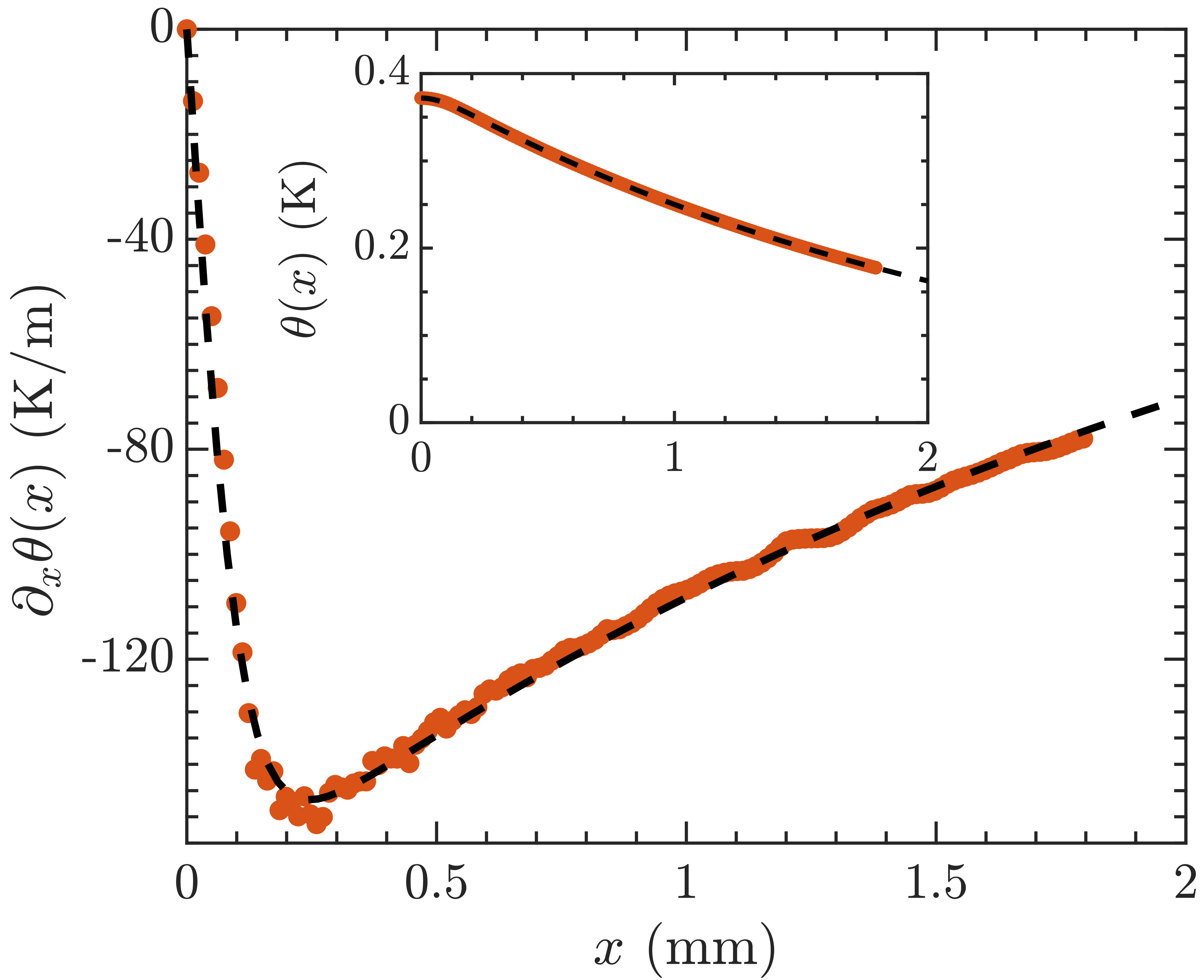}
	\caption{\label{fig:thetax}Derivative of the thermal field $\partial_x\theta(x)$ obtained experimentally from the collapse of curves shown in panel b) of Fig.~\ref{fig:master_curve_silicone} using equation (\ref{eq:theta_x}). The dashed line represents a fit to the experimentally determined gradient following the derivative of the thermal excess given by equation (\ref{eq:theta_heuristic}) with the parameter values given in the text. Inset: integrated thermal excess inferred from the experimentally determined gradient. The dashed line represents equation (\ref{eq:theta_heuristic}).}
\end{figure}

Following equation (\ref{eq:mastercurve}), the obtained master curve depends on the liquid properties, which are known, and on the thermal gradient driving the dynamics. The latter can hence be determined from the experimental master curve by inverting the relation given by (\ref{eq:mastercurve}) to get
\begin{equation}
	\partial_x\theta(x) 
	= 
	-\dfrac{2\eta}{\gamma_\theta}\dfrac{1}{\left\langle t \times h(x,t)\right\rangle^2_{exp}}\int_{0}^{x}\left\langle t \times h(s,t)\right\rangle_{exp}ds.
	\label{eq:theta_x}
\end{equation}
where $\left\langle t \times h(x,t)\right\rangle_{exp}$ is the experimentally determined master curve from the collapse of profiles as in figure~\ref{fig:master_curve_silicone}b), and is only a function of $x$.
Figure~\ref{fig:thetax} shows the thermal gradient $\partial_x\theta(x)$ obtained in this way.
Since we do not have data up to $x=0$, we extrapolate $\left\langle t \times h(x,t)\right\rangle_{exp}$ by a constant  for $x \leq \SI{0.1}{\milli\metre}$ in order to numerically compute the integral in equation~(\ref{eq:theta_x}). This extrapolation satisfies the symmetry requirement $\partial_xh(t,0) = 0$ and we have checked that it results in a negligible correction for larger values of $x$. 
A heuristic expression for $\partial_x\theta(x)$ can further be determined from the temperature gradient found from the master curve. Once integrated, it corresponds to a thermal field given by
\begin{equation} 
\theta(x) = \theta_{max}\left[\operatorname{sech}\left(\dfrac{x}{\delta}\right)\right]^{\delta/w}
\label{eq:theta_heuristic}
\end{equation}
where $\delta$ and $w$ are two characteristic lengths. The length $\delta$ accounts for the finite length of the zone of maximum temperature. This length results from both the finite size of the central line of the circuit dissipating heat and the diffusion of heat through the thickness of the substrate. Since the line width and substrate thickness are both of \SI{100}{\micro\metre}, $\delta$ is expected to be close to this value. The length $w$ is the decay length of the thermal field at distances from the central line larger than $\delta$ and smaller than roughly \SI{1}{cm}. We emphasise that at a longer range, the thermal excess is expected to go to zero asymptotically, with a characteristic decay length one order of magnitude larger than $w$. However, since we are interested in the central area of the film, for which submicron thicknesses are reached, equation (\ref{eq:theta_heuristic}) provides a satisfactory description of the thermal gradient in this area and further detail on the thermal field is not needed. In the particular case of Fig.~\ref{eq:theta_x}, a fit of equation (\ref{eq:theta_heuristic}) to the experimentally determined gradient yields the following values: $\theta_{max} = \SI{0.37}{\kelvin}$, $\delta=\SI{110}{\micro\metre}$ and $w =\SI{2.3}{\milli\metre}$.
These values are consistent with independent measurements of the thermal field at the bare substrate measured with an IR camera. We emphasise that the modest value of $\theta_{max}$ confirms that the liquids can be considered as having constant physical properties (i.e. viscosity, optical index and density), although they are heated.

In summary, the master evolution followed by the film thickness within intermediate times allows the determination of the thermal gradient. The empirical expression that is obtained can be used in numerical resolutions to test the consistency of the results, as described in the next section.

\subsection{Comparison of experimental data with numerical resolution \label{sb:TFE}}
\begin{figure}[h!]
	\centering
	\includegraphics[width=\linewidth]{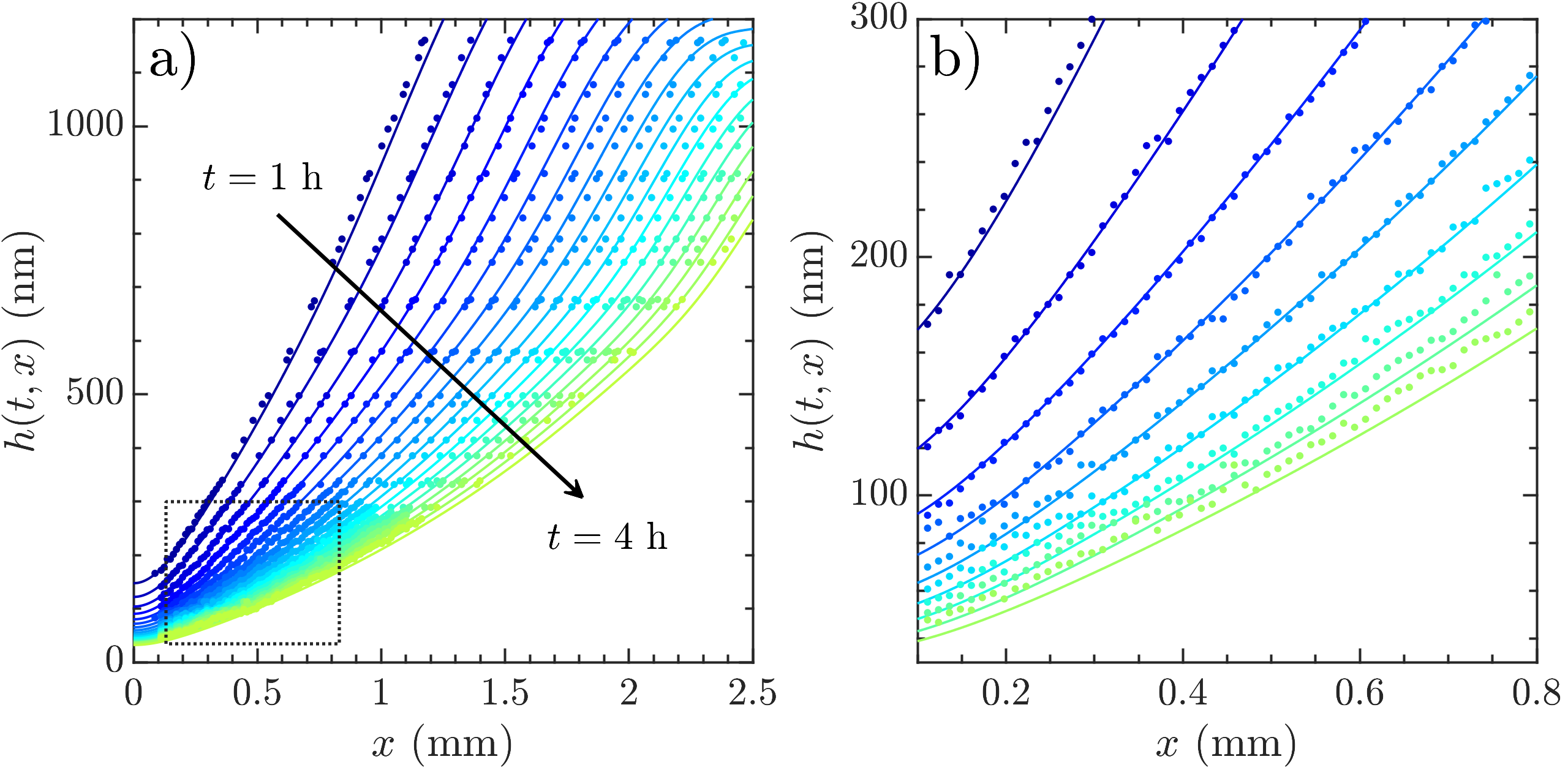}
	\caption{\label{fig:profiles_exp_and_num}a) Same experimental thickness profiles as in figure \ref{fig:master_curve_silicone}. The solid lines were obtained by numerical integration of equation (\ref{eq:TFE}), with the thermal gradient of figure \ref{fig:thetax} and a zero disjoining pressure.
	b) Zoomed view of the curves at small thicknesses showing the discrepancy between numerical and experimental data at this small scale (for better visualization, only one out of two curves is shown).}
\end{figure}

Since the thermal field is determined, equation (\ref{eq:TFE}) that includes capillary and gravity terms can be solved, allowing the comparison of experimental data with numerical results obtained without any assumption. We emphasise that, in the following analysis, we do not consider the disjoining pressure term since we focus here on large thickness values, for which intermolecular forces are negligible. In Fig.~\ref{fig:profiles_exp_and_num}, we can see that the numerical integration faithfully represents the measured local dynamics, even for thicknesses well below the micron scale. Therefore, the determination of the thermal gradient based on the assumption of negligible effects of capillarity and gravity is fully validated.

As expected, at thicknesses below $100$~nm, we observe that the numerical profiles deviate from experimental data.
In this thickness range, both capillarity and gravity can still be neglected and the dynamics results from thermocapillary and intermolecular effects. We have not tried to quantitatively describe the disjoining pressure term in silicone oil. Instead, we focus on a qualitative effect resulting from intermolecular forces that we have evidenced. 
	
\subsection{Transition to the ultra-thin film \label{sb:disjoining}}

We now examine the long term regime, when the thickness is on the order of tens of nanometres, and when
the effect of intermolecular forces becomes important.
In this regime, the reduced thermocapillary evolution (\ref{eq:hTC}) is not valid any more, and the curves $t\times h(t,x)$ are expected to deviate from the master curve.

\begin{figure}[h!]
	\centering
	\includegraphics[width=\linewidth]{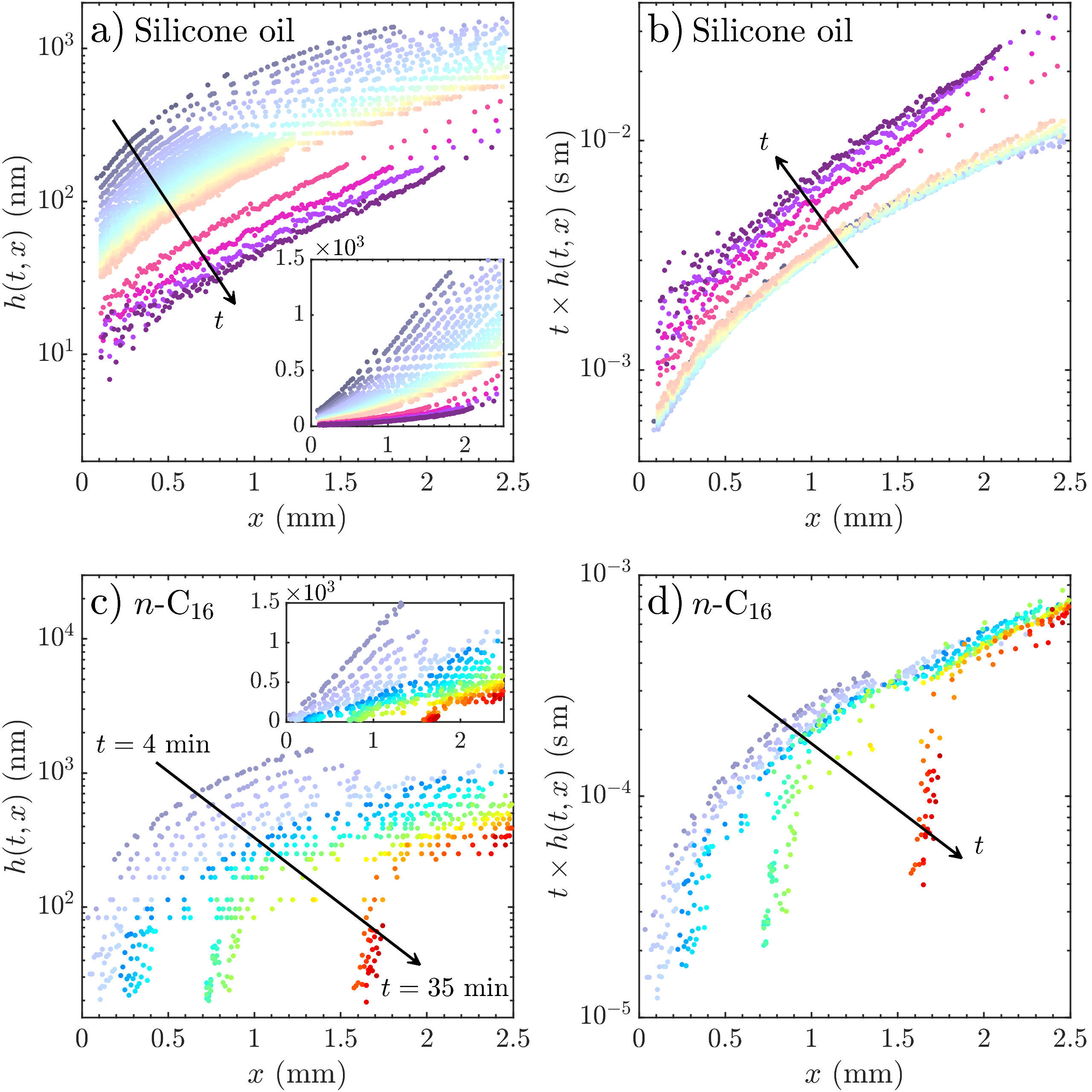}
	\caption{\label{fig:ht_longtimes}a) and b): same plots as in Fig.~\ref{fig:master_curve_silicone} (dimmed curves), with the addition of profiles at later times $t = 13$, $23$, $33$, \SI{43}{\hour}. The deviation of these last profiles from the master curve in panel b) reveals the effect of intermolecular forces.
	c) Experimental profiles $h(t,x)$, and d) curves $t \times h(t,x)$ for an experiment using n-hexadecane ($n$-C$_{16}$), with $h_0 = \SI{13.0\pm0.2}{\micro\metre}$ and $P = \SI{0.95\pm0.09}{\milli\watt}$. The time between profiles is \SI{1}{\minute}. Observe that the time arrow on panel d) points in the opposite direction as the corresponding arrow on panel b).}
\end{figure}

Figure~\ref{fig:ht_longtimes} shows the long-term regime of the evolution when the thickness enters the range in which intermolecular forces are relevant, for the same experiment as Fig.~\ref{fig:profiles_exp_and_num}.
The corresponding $t\times h(t,x)$ curves clearly lie above the master curve, which demonstrates that the thinning induced by thermocapillarity has been slowed down.
This deceleration can be explained by the presence of an effective repulsive intermolecular interaction that opposes the thermocapillary effect. A repulsion between the interfaces of the silicone oil film is expected since silicone oil totally wets the glass surface \cite{degennes_gouttes_2015}. Therefore, the thinning slows down and the final stationary state would correspond to an equilibrium of disjoining pressure and thermocapillary stress.

We have performed similar experiments with alkanes, and panels c) and d) of figure \ref{fig:ht_longtimes} display the thickness profiles measured in a film of $n$-hexadecane. Because of the smaller viscosity of $n$-hexadecane, the time scales of the experiment are much shorter. The curves at earlier times are nevertheless qualitatively similar to those of silicone oil and collapse as well onto a master curve. In contrast, at later times, the experimental profiles obtained with $n$-hexadecane reveal an acceleration of film thinning since they lie below the master curve. The acceleration is associated with an unstable behavior: the thickness profiles of $n$-hexadecane clearly  collapse and form a foot receding from the centre. We were however unable to evidence the presence or absence of an ultra-thin film in the central region, since, if it exists, its thickness lies below the sensitivity level of the measurement ($\SI{10}{\nano\metre}$). We can nevertheless state that, in the range of investigated thicknesses, the disjoining pressure for $n$-hexadecane is attractive since it accelerates thinning instead of opposing it, in contrast to silicone oil. 

This result is in contradiction with reports in the literature of hexadecane fully wetting glass surfaces \cite{shiri_thermal_2021}, hence suggesting that the interactions between the interfaces of $n$-hexadecane films should be repulsive. The attractive interactions we evidence rather evoke a partial or pseudo-partial wetting of $n$-hexadecane on glass. In the case of partial wetting, the spreading parameter is negative and the equilibrium situation is a drop sitting on the substrate with a finite angle of the contact line; in the case of pseudo-partial wetting, the spreading parameter is postive and the equilibrium situation is also a drop with a finite angle but that sits on a film of nanometric thickness. The latter situation results from non-monotonic variations of the disjoining pressure with thickness \cite{degennes_gouttes_2015}. Since glass is a material of high surface energy, it is likely that the spreading parameter is positive, favoring the hypothesis of pseudo-partial wetting.

We emphasise that we observe the same unexpected behavior with the ramified alkane HMN as with the linear $n$-hexadecane, and hence that it does not result from crystallisation effects. Actually, it is well known that surface freezing occurs in alkanes at temperatures up to \SI{3}{\kelvin} above their solidification temperature \cite{wu_surface_1993}, which affect the wetting behavior of $n$-hexadecane films \cite{lazar_spreading_2005-1, lu_stability_2010-1}. However, HMN has a crystallisation temperature which is well below room temperature, and remains liquid at room temperature even when present on sub-nanometric layers \cite{gosvami_squeeze-out_2008}. As a result, surface freezing cannot be invoked to explain the observed behaviors that are common to $n$-hexadecane and HMN. 
We rather suggest that these alkanes only pseudo-partially wet glass. In this case, the distinction from total wetting is difficult in macroscopic experiments if the equilibrium angles are small. In addition, times to attain an equilibrium state may reach the scale of the week \cite{perez_spreading_2001}. These difficulties are likely to explain why $n$-hexadecane has been suggested to totally wet glass, and show that static experiments at the macroscopic scale may not be sufficient to determine wetting behaviors properly.

\begin{figure}[h!]
	\centering
	\includegraphics[width=\linewidth]{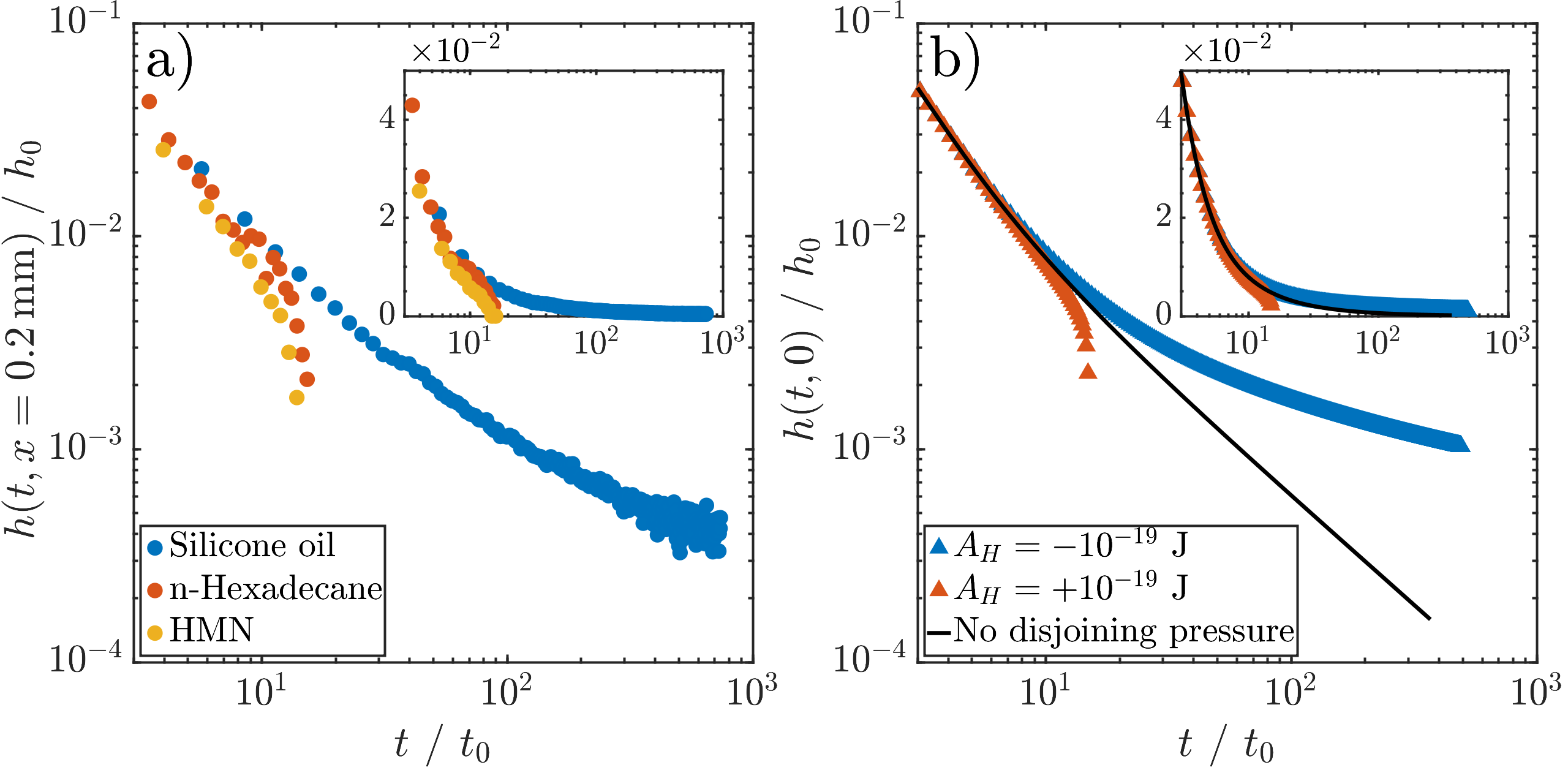}
	\caption{\label{fig:comparison_h_centre}a) Comparison of the thickness evolution near the centre, $h_c(t) = h(t, x = \SI{0.2}{\milli\metre})$, for the experiment with silicone oil and with $n$-hexadecane of Fig.~\ref{fig:ht_longtimes}.
	The time has been scaled by a thermocapillary timescale $t_0 =  2\eta w^2/(\Delta\gamma h_0)$ and the thickness by $h_0$. 
	Silicone oil: $h_0 = \SI{36\pm2}{\micro\metre}$, $\theta_{max} = \SI{0.37}{K}$, $w = \SI{2.3}{\milli\metre}$ ($t_0 \simeq \SI{211}{\second}$).
	n-hexadecane: $h_0 = \SI{13.0\pm0.2}{\micro\metre}$, $\theta_{max} = \SI{0.44}{K}$, $w = \SI{2.2}{\milli\metre}$ ($t_0 \simeq \SI{42}{\second}$).
	HMN: $h_0 = \SI{19.5\pm0.6}{\micro\metre}$, $\theta_{max} = \SI{0.54}{K}$, $w = \SI{2.2}{\milli\metre}$ ($t_0 \simeq \SI{30}{\second}$).
	b) Same as in a) but with profiles at the centre $h(t, 0)$ from the numerical integration of equation (\ref{eq:TFE}) with physical properties of $n$-hexadecane, when the intermolecular forces generate an effective repulsion ($A_H < 0$) vs. attraction ($A_H > 0$) of the interfaces. In the case where there is no disjoining pressure (e.g. $A_H = 0$), the evolution follows equation (\ref{eq:hTC}), i.e. $t \times h(t,0) \propto const.$ for $t \gg t_0$.}
\end{figure}

Additional insight is provided by numerical integration of the thin-film equation with a given disjoining pressure. In the apolar liquids and the range of film thicknesses we investigate, intermolecular forces are expected to be non-retarded van der Waals interactions $\Pi(h) = - A_H/(6\pi h^3)$, where $A_H$ is the Hamaker constant \cite{churaev_surface_2003}.
A repulsive interaction then corresponds to a negative Hamaker constant and an attraction to a positive Hamaker constant.
Using typical values for the thermal gradient, we have solved equation (\ref{eq:TFE}) with disjoining pressure terms of respectively attractive and repulsive nature. Because of numerical instabilities, full spatial variations as in figure \ref{fig:ht_longtimes} cannot be obtained in the case of attractive intermolecular forces. We nonetheless show in figure~(\ref{fig:comparison_h_centre}) the dynamics at the centre, $h(t,0)$, for attractive ($A_H > 0$) and repulsive ($A_H < 0$) interactions, as well as for zero disjoining pressure.
Since the dynamics at the centre cannot be measured in the experiment, the dynamics measured at a distance of $0.2$~mm from the centre $h(t, x = \SI{0.2}{\milli\metre})$ is displayed for experiments with silicone oil and both $n$-hexadecane and HMN. In order to compare the dynamics measured in different experimental conditions with liquids that have different viscosities, we have represented the dimensionless thickness a function of the ratio of the time and a thermocapillary timescale, $t_0 =  2\eta w^2/(\Delta\gamma h_0)$. As expected, the case $A_H = 0$ results in a pure-thermocapillary evolution $h(t,0) \propto t^{-1}$.
In contrast, for non-zero interactions, a deviation from $t^{-1}$ variations appears for thicknesses in the range of tens on nanometres. For $A_H < 0$ the thinning is slowed down, whereas for $A_H > 0$ it is abruptly accelerated, leading to film rupture.

Changing the sign of the disjoning pressure in the equation therefore satisfactorily accounts for the different behaviors observed. This result confirms that attractive intermolecular forces, at least in the thickness range investigated, are at stake between the interfaces of films of the considered alkanes. Beyond this qualitative finding, the very similar shape of the experimental and computed curves suggest that the expression for disjoining pressure we use is close to the real one. However, we have found that the dynamics in silicone oil could not be quantitatively described by the simple law we have used: clearly in figure~(\ref{fig:comparison_h_centre}), the experimental thinning is faster than in the computation. Consistently, the full thickness profiles, as the ones reported in figure~\ref{fig:profiles_exp_and_num} could not be described, even when changing the value of the Hamaker constant. This discrepancy may result from the fact that the used silicone oil is not monodisperse but rather constituted by a mixture of macromolecules of different lengths. Mixtures effects have actually been predicted to change the exponent of the thickness-dependency of molecular interactions \cite{derjaguin_question_1978}. These complex effects are out of the scope of the present work and we have not tried to further determine the disjoining pressure at stake. However, since silicone oil is widely used as a model wetting liquid, it could be an interesting aim for future research.

\section{Conclusion \label{s:conclusion}}
In conclusion, we have demonstrated that the dynamics of thin films induced by a thermal gradient can serve as a tool for investigating intermolecular forces. We predict and experimentally evidence a regime in which thermocapillarity fully drives thinning. The thermal gradient can thus be determined from thickness profile measurements and, in turn, be used to compute the expected dynamics when intermolecular forces are at stake. We demonstrate that thickness measurements allow a clear discrimination between the cases of attractive and repulsive interactions between the interfaces of the liquid films. The experiments reveal an unexpected behavior of both linear and ramified alkanes when film thicknesses reach a few tens of nanometres, and thus show that accurate probes of liquid films at very small scales are still needed.

\begin{acknowledgements}
This project has received funding from the European Union’s Horizon 2020 research and innovation program under the Marie Sk\l{}odowska-Curie grant agreement No 754387.
\end{acknowledgements}

\bibliography{bibliography.bib}

\end{document}